\documentclass{ifacconf}

\usepackage{graphicx}      
\usepackage{natbib}        
\usepackage{amsmath}
\begin{document}
\begin{frontmatter}

\title{Increasing Traffic Throughput by Controlling Autonomous Vehicles at Low Penetration Rates \thanksref{footnoteinfo}} 

\thanks[footnoteinfo]{This work was supported in part by the National Science Foundation project CMMI-1462289, the Natural Science Foundation of China (NSFC) project \# 71428001, the US DOT Center for Transportation, Environment, and Community Health (CTECH), and the Lloyd’s Register Foundation, UK. }

\author[First]{Ronan L. Keane} 
\author[Second]{H. Oliver Gao} 

\address[First]{Systems Engineering, Cornell University, 
   Ithaca, NY (email: rlk268@cornell.edu).}
\address[Second]{Civil and Environmental Engineering, Cornell University, 
   Ithaca, NY (email: hg55@cornell.edu).}

\begin{abstract}                
Human drivers may behave in an imprecise/unstable manner, leading to traffic oscillations which are harmful to traffic throughput. Recent field experiments have shown that the control of a single autonomous vehicle (AV) can increase traffic throughput on a circular test track, as well as reduce traffic oscillations on straight roads. We consider a mixed traffic environment consisting of humans and autonomous vehicles, where the goal is to find a control policy for the autonomous vehicles which maximizes traffic throughput by preventing oscillations in speed. We formulate this problem as an optimization problem which can be solved using gradient based optimization. Numerical experiments on a circular road show that the optimized control policy improves traffic throughput by 28\%.
\end{abstract}


\end{frontmatter}
\section{Introduction}
As the adoption of autonomous vehicles (AV) becomes ever closer to becoming a reality, many questions have arisen in regards to how their usage will shape the future of mobility. The potential effects AV will have on energy usage, air pollution, mode choice, and traffic congestion are both profound and uncertain. In this paper we consider the possible effects that a small number of AV can have on vehicular traffic flow. We consider traffic simulations consisting of a mix of human and autonomous vehicles, where the autonomous vehicles are present at low penetration rates ($<$5\% of total vehicles). We assume that the behavior of the AV can be controlled through a decentralized strategy, where all AV only have access to local information and follow some control policy which is the same for all AV. The human-driven vehicles cannot be directly controlled and instead follow a realistic model of human driving behavior. Our goal is to implement a control policy for the AV which maximizes some performance metric such as average speed in the simulation. 
\section{Background}
\subsection{Microscopic Traffic Flow Theory}
In general, traffic flow models can be categorized into three broad categories: fluid-based models, queueing based models, and agent-based models. In this work we are concerned with the agent based models, also referred to as microscopic models. Of particular interest are ``car-following'' models, which describe the longitudinal behavior of vehicles (the evolution of the position of a vehicle, i.e. the vehicle's trajectory). Car following models are based on the idea that your trajectory is largely dependent on that of your lead vehicle, the vehicle directly in front of you. One such commonly used car following model is the Intelligent Driver Model (IDM)
\begin{align*} 
\ddot x(t) = c_4\left( 1 - \left( \dfrac{\dot x(t)}{c_1} \right)^4  - \left( \dfrac{s^*}{s(t)} \right)^2\right) \tag{1}\label{IDM}\\
s^*  = c_3+c_2\dot x(t) + \dfrac{\dot x(t) (\dot x(t) - \dot x_{l}(t))}{2 \sqrt{c_4 c_5}}
\end{align*}
Where $x$ is the position of the vehicle being modeled (so that $\dot x$ and $\ddot x$ are the speed and acceleration), $x_l$ is the position of the lead vehicle, $s = x_l - x $ is the space headway (distance to the vehicle in front of you), and $c_1 - c_5$ are model parameters. \\
 A vehicle can be said to be in equilibrium if $\ddot x = \ddot x_l = 0$ and $\dot x = \dot x_l$, that is if vehicle $x$ and its leader are both maintaining a constant speed. For any particular speed $v$, there is a headway $s_{eq}$ such that if $\dot x = \dot x_l = v$ and $s = s_{eq}$ then $\ddot x = 0$. Then for a chain of vehicles all with identical parameters, they are said to be in the equilibrium solution if all vehicles have speed $v$ and headway $s_{eq}$. \\
Besides the equilibrium solution, another vital concept of car following models is string stability. Consider an oscillatory lead vehicle, meaning that instead of $\dot x_l(t)$ being constant, $\dot x_l(t)$ has a changing speed. A model is said to be string stable if the oscillation present in $\dot x(t)$ is less severe than in $\dot x_l(t)$. This means that as the oscillation passes through a long chain of vehicles, it will gradually disappear. 
\subsection{Control of Autonomous Vehicles}
The equilibrium solution corresponds to the highest possible flow state of the vehicles (see \cite{article}). Thus a natural goal of traffic control is to try to \textit{homogenize} traffic, that is to try to ensure that all vehicles are travelling at approximately the same speed inside the congested regions of traffic. One way of accomplishing this is to try to ensure traffic is string stable, so that perturbations which occur naturally in traffic (for example, from lane changing or random human error) will not gradually amplify into larger oscillations. In \cite{FS} a field experiment was performed with actual human drivers and a single controlled vehicle on a circular test track. In that experiment, with no control, an oscillatory state emerged on the track, resulting in an average throughput of 1827 veh/hr (vehicles per hour). By controlling a single AV, traffic on the test track is stabilized and a throughput of 2085 veh/hr is achieved. The AV in that experiment used a nonlinear controller called follower stopper. \\
In \cite{linearcav} field experiments were performed on straight roads with a similar setup consisting of several human drivers with a single AV. In that paper they showed humans drove in an unstable way which caused oscillations in earlier vehicles to be amplified in later ones. When the AV reacted to the oscillation however, the oscillation decreased in amplitude and additionally resulted in the following human-driven vehicles amplifying the oscillations by a smaller amount. That experiment used a custom linear controller for the AV which had the ability to take into account information from multiple vehicles (and not just the immediate leader). \\
In this paper, we consider a decentralized control strategy for AV which results in an increased throughput by causing traffic to be string stable. 

\section{Methodology}
\subsection{Problem Formulation}
We formulate the control policy for the AV as the solution to the following optimization problem
\begin{align*} 
& \underset{p}{\min} \quad F = \sum_{t = t_0}^{t_1} l(s_t) \tag{2}\label{opt} \\ 
& s.t. \quad s_{t+1} = u(s_t, a_t) \ \ \ t \in [t_0, t_1 - 1]\\
& \quad \quad  \  a_t = f(p, p_h, s_t) \ \ \ t \in [t_0, t_1 - 1]
\end{align*}
Subscripts indicate time index. $s_t$ is the state, it consists of the position, speed, and space headway for each vehicle in the simulation. $a_t$ are the actions, which consist of the acceleration for each vehicle in the simulation. $l$ is a loss function which represents the cost incurred at the current state $s_t$.  The objective $F$ is to minimize the total cost incurred over the total simulation. $f$ maps from the current state (position, speed, headway for each vehicle) to the action taken (acceleration each vehicle takes at the current timestep). $f$ consists of both the parameters for the AV control policy (denoted $p$) as well as parameters which control how the human-driven vehicles behave (denoted $p_h$). $u$ is a function which updates the current state based on the current actions. \\
In this paper $f$ consists of microscopic traffic models (car following models) for the humans and a parametrized control policy for the AV. $u$ is a first-order forward euler scheme. \\
The loss function used is 
\begin{align*} 
l(s_t) = \sum_{i = 1}^n (v^i - v^i_{\rm avg})^2 - v^i_{\rm avg} + \mathcal{P}
\end{align*}
Superscripts indicate vehicle index (time dependence suppressed for clarity). $v^i$ is the current speed of vehicle $i$ and $v_{\rm avg}^i$ is the average speed of vehicle $i$. The first term encourages vehicle speeds to be close to the average, i.e. it penalizes oscillatory speeds which correspond to traffic congestion. The second term encourages the vehicle speeds to be as large as possible. The last term is a penalty which grows exponentially large if the space headway becomes too small. It is zero under normal circumstances and serves to prevent simulations which result in vehicle collisions.

\subsection{Differentiation of the Optimization Problem}
The gradient of Eq. \eqref{opt} can be found efficiently using reverse mode differentiation. Using automatic differentiation naively on Eq. \eqref{opt} leads to difficulties when the number of timesteps in the simulation is large. This is because the size of a computational graph is proportional to the number of timesteps and vehicles in the simulation; this means a sufficiently complex simulation results in a massive computational graph which becomes intractable for automatic differentiation programs. A solution to this problem is given in \cite{neuralode}, where they combine (reverse mode) automatic differentiation with the adjoint method \cite{adjoint}. Specifically, the technique used is to write down the adjoint equations by hand, and use automatic differentiation to calculate all the partial derivatives in the resulting derivation. This results in a gradient calculation that is both computationally tractable as well as significantly less tedious to implement than the adjoint method. We give the steps required to calculate the gradient but omit the derivation. 
\begin{enumerate}
\item Compute $F$ as in \eqref{opt}, keep all states/actions $s_t, a_t$ in memory. 
\item set $\lambda_{t_1-1} = \dfrac{\partial l (s_{t_1})}{\partial s_t}$
\item For times $t \in [t_1 - 1, t_0]$: \\
set $\mu_t = \lambda_t  \frac{\partial u}{\partial a_t}$ \\
set $\lambda_{t-1} = \lambda_t\frac{\partial u}{\partial s_t} + \mu_t \frac{\partial f}{\partial s_t} - \frac{\partial l}{\partial s_t}$
\item Compute gradient $\frac{\partial F}{\partial p}  = \sum_{t = t_0}^{t_1-1} -\mu_t \frac{\partial f}{\partial p}$
\end{enumerate}

 \cite{fastcalibration} treats Eq. \eqref{opt} written in a time-continuous form, and shows that the gradient is continuous under the assumptions that all the partial derivatives $\frac{\partial l}{\partial s_t}, \frac{\partial f}{\partial p}, \frac{\partial u}{\partial a_t}, \frac{\partial u}{\partial s_t}, \frac{\partial f}{\partial s_t}$ are \textit{piecewise} uniformly Lipschitz continuous. 
 
\section{Numerical Experiment}
\subsection{Experiment Setup}
We take the human-driven vehicles in the simulation to be controlled by the IDM (Eq. \eqref{IDM}) with parameters $[c_1 - c_5] = [33.33, 1.2, 2, 1.1, 1.5]$. We initially consider a circular road where all vehicles start in equilibrium and are controlled by humans. A perturbation is applied to the simulation so that after some number of timesteps the vehicles will be in an oscillatory state with reduced throughput. Then starting from this reduced flow state, a control policy is applied to a single vehicle (the AV). We consider as our class of parametrized policies, controllers proposed by different works in the literature: the follower-stopper (FS) from \cite{FS} and the linear control model from \cite{linearcav}. We also consider using IDM as a controller, even though it was designed to replicate human driving and not for control. The parameters of the controller are optimized by solving Eq. \eqref{opt} using the gradient based convex optimization algorithm l-bfgs-b described in \cite{lbfgsb}.  
\subsection{Results}
The length of the test track is choosen so that the equilibrium solution corresponds to a speed of 15 m/s, so the maximum flow state is achieved when all vehicles are traveling at 15 m/s. The situation with no control is shown in Fig. 1. Traffic is observed in a highly oscillatory state with a pronounced stop and go wave, causing vehicles to oscillate between 0 and 25 m/s. The average speed in the simulation is only 11.38 m/s. The results of applying the control are summarized in table 1. 
\begin{table}[hb]
\begin{center}
\caption{Average speed measured over a 20 minute time period starting from the oscillatory state.}\label{table1}
\begin{tabular}{ccccc}
 & No Control & FS & Linear Control & IDM \\\hline
Avg. Speed (m/s) & 11.38 & 14.52& 14.41 &  14.47\\ 
\end{tabular}
\end{center}
\end{table}
After applying any of the optimized control strategies, the average speed is  greatly increased and traffic returns to the maximum flow state after approximately 1200 timesteps (5 minutes). Fig 2 shows the simulation with a single AV being controlled using the default follower stopper (FS) parameters. The controller creates gaps in traffic which allow the oscillation to dissipate, but the gaps being created are not large enough to fully dissipate the oscillation, so it takes traffic a large time to stabilize. The average speed with the default parameters is only 14.06 m/s. Fig. 3 shows the situation after optimizing the parameters; in this case traffic is stabilized much faster and the average speed is increased to 14.52 m/s.

\begin{figure}
\begin{center}
\includegraphics[width=8.4cm]{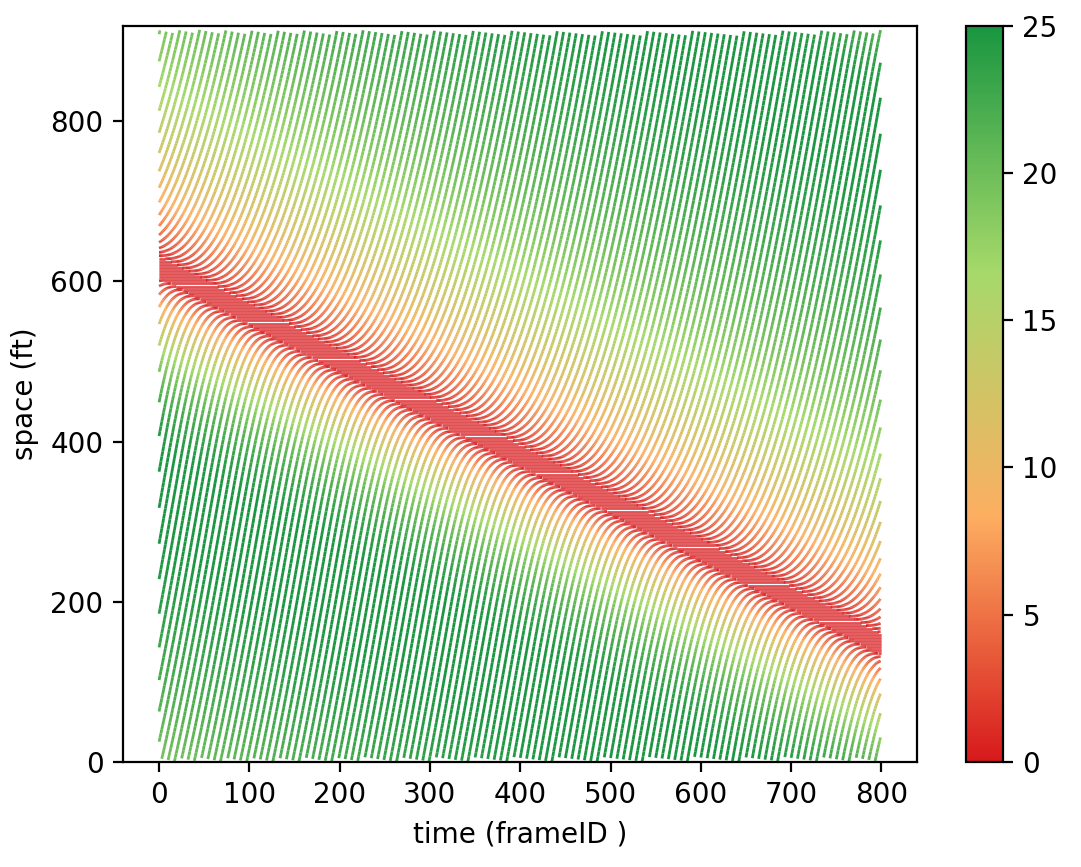}    
\caption{Plot of trajectories in the baseline scenario (no control). Trajectories are colored based on their speed (between 0-25 m/s). X axis is measured in timesteps (.25 s).} 
\label{}
\end{center}
\end{figure}

\begin{figure}
\begin{center}
\includegraphics[width=8.4cm]{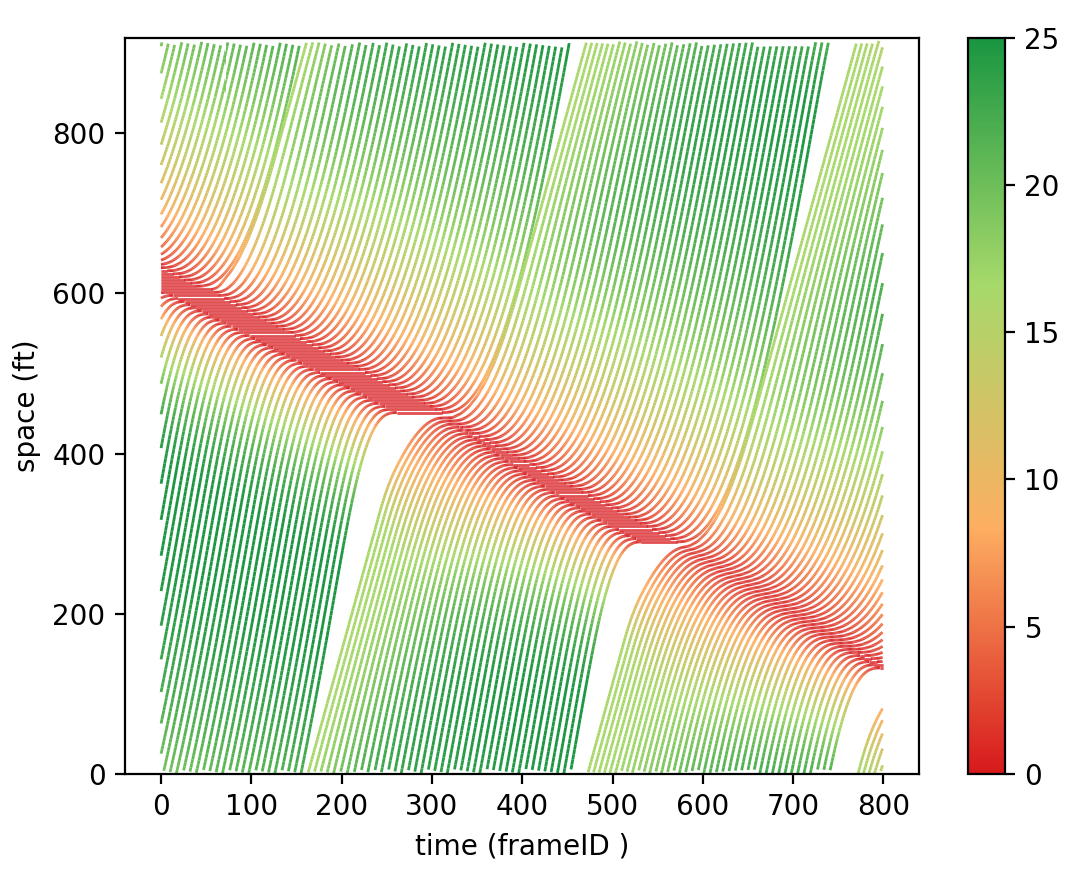}    
\caption{Plot of trajectories with a single AV being controlled, using the default control parameters.} 
\label{fig:bifurcation}
\end{center}
\end{figure}

\begin{figure}
\begin{center}
\includegraphics[width=8.4cm]{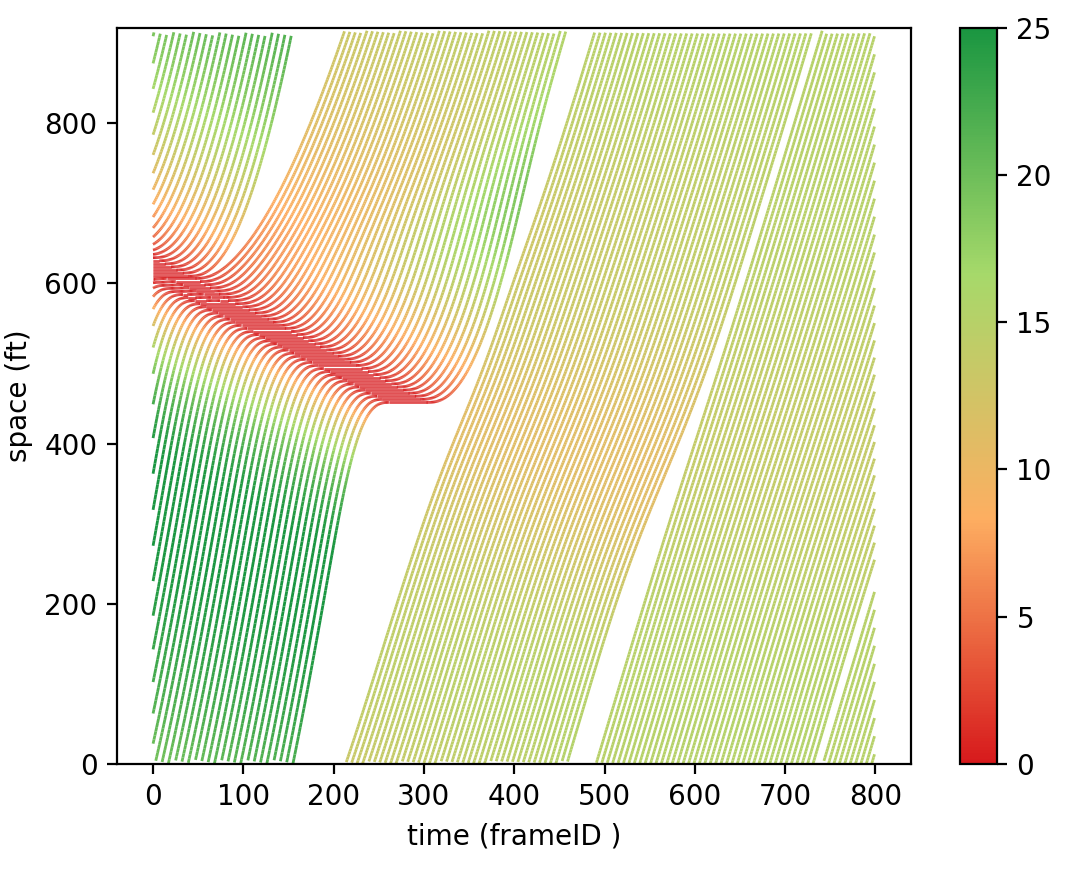}    
\caption{Plot of trajectories with a single AV being controlled with optimal parameters for follower stopper. The AV creates a large gap in the traffic which stops the oscillation from propagating and allows traffic to return to an equilibrium state.} 
\label{fig:bifurcation}
\end{center}
\end{figure}

\section{Conclusion}
In this paper we considered the control of a single autonomous vehicle (AV) on a circular test track with human-driven vehicles. The AV followed a parametrized control policy with the parameters found by using a gradient-based convex optimization algorithm. In our experiments, we showed that a single AV can stabilize traffic flow and result in a 28\% increase in average speeds compared to the case of no control.

\bibliography{ifacconf}             
                                                   







\end{document}